\documentclass[conference,10pt]{IEEEtran}
\usepackage{epsfig,rotating,setspace,latexsym,amsmath,epsf,amssymb,amsfonts,bm,theorem,subfigure,epstopdf}
\usepackage{cite,authblk}
\usepackage{bbm}
\usepackage{algorithmic,algorithm}

\newtheorem{lemma}{Lemma}
\newenvironment{Proof}[1]{\medskip\par\noindent{\bf Proof:\,}\,#1}{{\mbox{\,$\blacksquare$}\par}}
\IEEEoverridecommandlockouts

\allowdisplaybreaks
\usepackage{color}
\usepackage{graphicx}
\usepackage[utf8]{inputenc}
\usepackage{amsmath}
\usepackage{calc}

\begin{document}
	
	\title{Age of Information with Gilbert-Elliot \\Servers and Samplers\thanks{This work was supported by NSF Grants CCF 17-13977 and ECCS 18-07348.}}

	\author{Baturalp Buyukates \qquad Sennur Ulukus\\
	\normalsize Department of Electrical and Computer Engineering\\
	\normalsize University of Maryland, College Park, MD 20742\\
	\normalsize  \emph{baturalp@umd.edu} \qquad \emph{ulukus@umd.edu}}

	\maketitle
	
\begin{abstract}	
We study age of information in a status updating system that consists of a single sampler, i.e., source node, that sends time-sensitive status updates to a single monitor node through a server node. We first consider a Gilbert-Elliot~service profile at the server node. In this model, service times at the server node follow a finite state Markov chain with two states: \emph{bad} state $b$ and \emph{good} state $g$ where the server is faster in state $g$. We determine the time average age experienced by the monitor node and characterize the age-optimal state transition matrix $P$ with and without an average cost constraint on the service operation. Next, we consider a Gilbert-Elliot sampling profile at the source. In this model, the interarrival times follow a finite state Markov chain with two states: \emph{bad} state $b$ and \emph{good} state $g$ where samples are more frequent in state $g$. We find the time average age experienced by the monitor node and characterize the age-optimal state transition matrix $P$.
\end{abstract}

\section{Introduction}

Age of information (AoI) is a network performance metric which has been proposed to assess the timeliness in real-time status updating systems. Such systems include sensor networks, vehicular networks, and emergency alarm systems. In all these systems, time-critical data generated by the source node(s) are sent to the interested monitor node(s). Here, the update packets are time-critical since the information loses its value as it becomes stale. Thus, overall freshness of the information is desired which motivates the study of age of information in communication networks with various foci, e.g., queueing theory \cite{Kaul12a, Costa16,Yates12,Najm17,Inoue18b, Soysal18, Soysal19, Zhong17a,Buyukates18, Buyukates19, Buyukates18b, Maatouk19, Tripathi19}, scheduling and energy harvesting \cite{Hsu17,Zhou18b, Bastopcu19b, Bastopcu19,Bastopcu20b, Bastopcu20c, Ioannidis09, Buyukates18c, Buyukates19b, Bacinoglu15,Wu18,Arafa18a,Arafa18d,Farazi18, Baknina18b, Leng19c}, and caching and coding \cite{Yates17b, Kam17b, Zhong16, Parag17, Sac18, Mayekar18, MelihBatu1, Arafa19b, Buyukates19c, Bastopcu20}.

In all these works, there is an underlying i.i.d.~structure in the system. The service times and packet interarrivals are i.i.d.~processes and the focus is on analyzing and optimizing the resulting age of information. There may be scenarios in which these processes are correlated over time or over different status update packets. Reference \cite{Sun17a} models transmission times as a stationary and ergodic Markov chain to analyze the effect of the temporal correlation between transmission times on age-optimal scheduling.  References \cite{Costa15b, Nguyen19} study information freshness over Markovian channels. Specifically, reference \cite{Costa15b} models the channel using a Gilbert-Elliot model and introduces the concept of channel information age. This metric is used to express the utility and analyze the effect of aging on the probability of error in estimating the channel state. 
Reference \cite{Kam18a} studies the freshness over a network with a Markov source and proposes an effective age metric which captures estimation error as well as timeliness. Reference \cite{Huang17} studies age of information with a two-state Markov modulated service process and characterizes the average age for an FCFS operation under infinite and zero buffer size settings. 

\begin{figure}[t]
	\centering  \includegraphics[width=1\columnwidth]{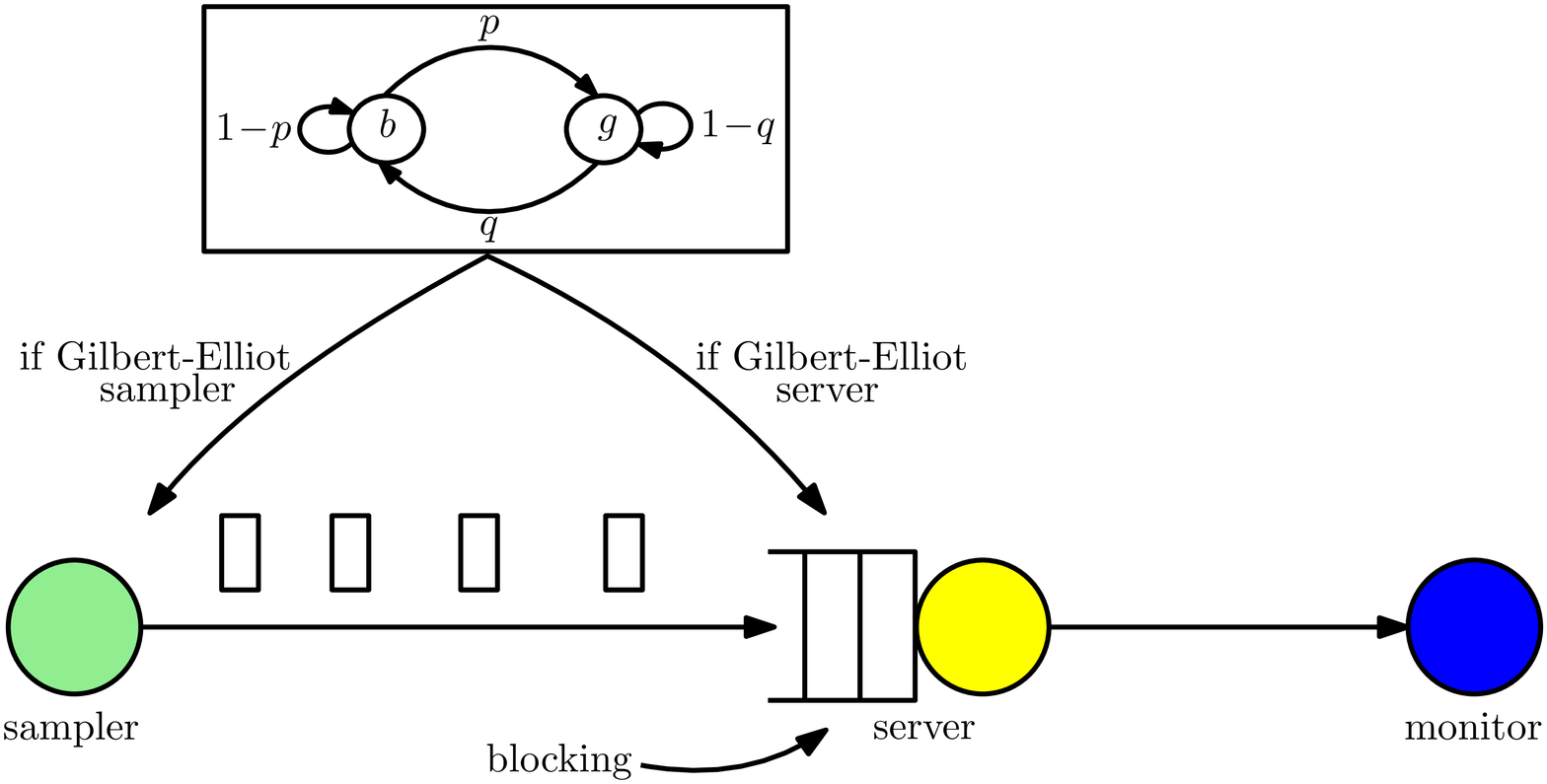}
	\caption{A single sampler sends time-sensitive status updates to a single monitor node through a server node. We consider Gilbert-Elliot server and Gilbert-Elliot sampler settings.}
	\label{fig:model}
	\vspace{-0.5cm}
\end{figure}

In this work, we consider a status updating system in which there is a single sampler which takes samples from an observed phenomenon and sends them to an interested monitor node in the form of status update packets through a single server node (see Fig.~\ref{fig:model}). We study age of information with Gilbert-Elliot servers and samplers under blocking packet management policy at the server node. We first analyze the case in which the service times follow a finite state Markov chain with two states: \emph{bad} state $b$ and \emph{good} state $g$ such that in state $g$, the service performance is faster than that in state $b$. The motivation for studying this kind of a service profile comes from the measurements over Amazon EC2 clusters which show high variability in computing speeds of the servers over time \cite{Yang19}. These measurements indicate that when a server is in a certain state it tends to stay in that state in the next rounds of the computation which implies a dependence in service times over time. In addition, \cite{Ploger19} shows that the channel quality and reliability in cooperative driving follows a Markov-modulated process which depends on the number of interfering vehicles thereby affecting the service performance. Further, in sensor networking applications, energy constraints on servers may prevent them from operating in faster states all the time, and may force them to switch between faster and slower states. 
 
We derive the time average age under this Markovian service profile and characterize the age-optimal state transition matrix of the underlying Markov chain first without considering any constraints on the operation of the system. Next, we consider a more realistic scenario in which each state has an operational cost and the system is subject to an overall budget. 

Next, we consider the case in which the sampler, i.e, the source node, follows a Gilbert-Elliot model based operation. The sampler takes samples based on a two-state Markov chain, as in Fig.~\ref{fig:model}. When in state $g$, it samples the observed phenomenon more frequently whereas in state $b$, samples are taken more sparsely. Thus, under this operation, interarrivals to the server node are no longer i.i.d. but follow a two-state Markov chain. This non i.i.d. sampling operation is particularly relevant when the sampler's operation cost is considered as the sampler may not be able to afford taking frequent samples all the time and may switch to a low-cost operation to save energy. Further, the sampler may choose to sample the process more often when the observed process varies above a certain threshold or varies too fast. We characterize the average age under such Gilbert-Elliot sampling and find the age-optimal state transition matrix $P$. 

\section{System Model and Age Metric} \label{model}

We consider a communication system (see Fig.~\ref{fig:model}), where there is a single sampler that 
takes samples from an observed phenomenon at random and immediately transmits these samples to a monitor node through a single server node in the form of status update packets. The server node implements a blocking policy in which an arriving update packet goes directly into service only if the server is idle. Update packets arriving when the server node is busy are discarded. 

Unlike most of the literature, we consider non i.i.d.~service and interarrival profiles. Here, we use a simple Gilbert-Elliot model to introduce a non i.i.d.~structure to the system. We consider two scenarios: Gilbert-Elliot service times and i.i.d.~interarrival times; and i.i.d.~service times and Gilbert-Elliot interarrival times.

When the server follows a Gilbert-Elliot model, service times $S$ follow a two state Markov chain with states \emph{bad} ($b$) and \emph{good} ($g$) such that in state $b$, the server node is slower and the service takes longer than the service in state $g$. We model the service times with exponential random variables $S_b$ with rate $\mu_b$ in state $b$ and $S_g$ with rate $\mu_g$ in state $g$ where $\mu_g > \mu_b$ as in \cite{Huang17}. In this case, we model the update arrivals at the server node as a Poisson process with rate $\lambda$. We adopt an event-triggered Markov chain in which the state change only occurs when a new packet enters service. Thus, during the service of an update packet, service performance remains the same. The transition probability from state $b$ to state $g$ is $p$ and the transition probability from state $g$ to state $b$ is $q$ where $p \in (0,1)$ and $q \in (0,1)$ (see Fig.~\ref{fig:model}). State transition matrix $P$ of this Markov chain is 
\begin{align}
P = \begin{bmatrix}
 1-p & p\\
 q & 1-q  \label{state_trans}
\end{bmatrix}.
\end{align}

We note that this Markov chain is irreducible, aperiodic, and positive recurrent. Thus, service times constitute an ergodic Markov chain with a stationary distribution,
\begin{align}
	P(S = S_b) = \frac{q}{p+q},\qquad P(S = S_g) = \frac{p}{p+q}, \label{stat_dist}
\end{align}
where $S$ is the service time of a packet that enters service. 

When the sampler follows a Gilbert-Elliot model, this time, update interarrivals at the server node constitute a Markov chain with the state transition matrix in (\ref{state_trans}) where in state $g$ interarrival times are exponential random variables with rate $\lambda_g$ and in state $b$ interarrival times are exponential random variables with rate $\lambda_b$ where $\lambda_g > \lambda_b$ to reflect the increased sampling frequency in state $g$. The Markov chain is again event-triggered such that the sampler's state changes whenever an update packet enters service at the server node.

To quantify the timeliness in the system, we use the age of information metric. At time $t$ age at the monitor node is the random process $\Delta(t) = t - u(t)$ where $u(t)$ is the time-stamp of the most recent update at the destination node. The metric we use, time averaged age, is given by
\begin{align}
\Delta = \lim_{\tau\to\infty} \frac{1}{\tau} \int_{0}^{\tau} \Delta(t) dt,
\end{align}
where $\Delta(t)$ is the instantaneous age at the monitor node. 

In the next section, we derive an average age expression for the cases of Gilbert-Elliot servers and samplers. 

\section{Average Age Analysis }\label{age_analysis}

The sampler generates status update packets and immediately sends them through a delay-free link to the server node. Since a blocking policy is implemented, only the packets that find the server idle go into service. We denote such packets that enter service at the server node as successful packets. Let $T_{j-1}$ denote the time at which the $j$th successful update packet is generated at the sampler. Since newly generated packets are assumed to be instantaneously available to the server node, $T_{j-1}$ also marks the time at which the $j$th successful update packet arrives at the server node. Random variable $Y$ denotes the update cycle at the server node, the time in between two successful arrivals, where $Y_j = T_j - T_{j-1}$.

\begin{figure}[t]
	\centering  \includegraphics[width=0.80\columnwidth]{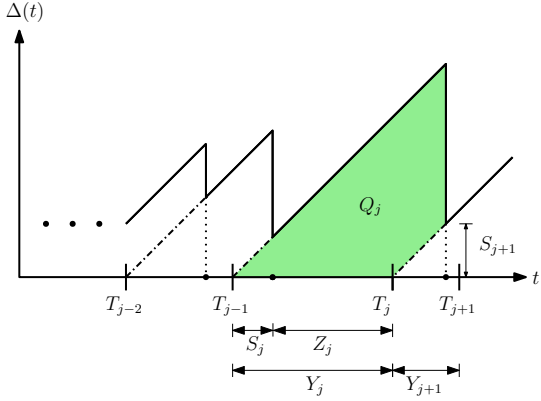}
	\caption{Sample age evolution $\Delta{(t)}$ at the monitor node. Successful updates are indexed by $j$. The $j$th successful update arrives at the server node at $T_{j-1}$. Update cycle at the server node is the time in between two successive arrivals and is equal to $Y_j = S_j + Z_j = T_{j} - T_{j-1}$.}
	\label{fig:ageEvol}
	\vspace{-0.5cm}
\end{figure}

Update cycle $Y_j$ consists of service time $S_j$ and idle waiting time $Z_j$ as the server needs to wait for the next arrival upon an update delivery to the monitor node such that
\begin{align}
Y_j = S_j + Z_j. \label{update_cycle}
\end{align}
We note that $S_j$ and $Z_j$ are mutually independent as the arrival and service processes are independent. Sample age evolution at the destination node is given in Fig.~\ref{fig:ageEvol}. Here, $Q_j$ denotes the area under the instantaneous age curve in update cycle $j$ and $Y_j$ denotes the length of the $j$th update cycle as defined earlier. The metric we use, long term average age, is the average area under the age curve which is given \cite{Kaul12a, Yates12} by 
\begin{align}
\Delta = \limsup_{n\to\infty} \frac{\frac{1}{n}\sum_{j=1}^{n} Q_j}{\frac{1}{n}\sum_{j=1}^{n}Y_j}. \label{avg_age1}
\end{align}

In the next two subsections, we find the average age for Gilbert-Elliot service times and interarrival times, respectively.

\subsection{Gilbert-Elliot Service Times and I.i.d.~Interarrival Times}\label{service}

Status update packets arrive at the server node as a Poisson process with rate $\lambda$. Thus, update packet interarrivals at the server node are i.i.d.~exponential random variables with rate $\lambda$. Due to the memoryless property of the update arrivals at the server node, $Z$ is also exponentially distributed with rate $\lambda$. Service times $S$ follow a two-state Markov chain. Thus, two consecutive service times are dependent through the state transition matrix $P$ given in (\ref{state_trans}). 

Conditioned on $S_j$ and $S_{j+1}$, the $j$th update cycle and the area under the age curve in this cycle are characterized by
\begin{align}
(Y_j, Q_j) =  \begin{cases} 
	(Y_b, Q_{bb}),  & \text{if } S_j = S_b,\quad S_{j+1} = S_b \\
	(Y_b, Q_{bg}),  & \text{if } S_j = S_b,\quad S_{j+1} = S_g \\
	(Y_g, Q_{gb}),  & \text{if } S_j = S_g,\quad S_{j+1} = S_b \\
	(Y_g, Q_{gg}),  & \text{if } S_j = S_g,\quad S_{j+1} = S_g.
	\end{cases} \label{updatecycle_pair}
\end{align} 

We note that when the service times and interarrival times are i.i.d., we have a renewal process with inter-renewal time equal to the update cycle $Y$. However, in our model, update cycles $Y_j$ do not form an i.i.d.~sequence unlike the prior models considered in the literature. Rather, each $(Y_j, Q_j)$ is characterized as in (\ref{updatecycle_pair}) depending on the state of the service time in update cycles $j$ and $j+1$. Since the underlying Markov chain is stationary and ergodic, we have $(Y_j, Q_j) \sim (Y, Q)$ over all update cycles and (\ref{avg_age1}) reduces to
\begin{align}
\Delta = \frac{E[Q]}{E[Y]} = \frac{qE[Q_b] + pE[Q_g]}{qE[Y_b] + pE[Y_g]}, \label{avg_age2}
\end{align}
where the first equality follows from \cite[Appendix A]{Sun17a} and the second equality follows from the law of total probability where we define $E[Q_b] \!=\! (1\!-\!p)E[Q_{bb}] + pE[Q_{bg}]$ and $E[Q_g] \!=\! qE[Q_{gb}] + (1\!-\!q)E[Q_{gg}]$. From this, along with Fig.~\ref{fig:ageEvol}, we get
\begin{align}
E [Q_b] =& \frac{1}{2}E[(S_b + Z)^2] + (E[S_b] + E[Z])E[\bar{S}], \label{eqn_third}\\
E [Q_g] =& \frac{1}{2}E[(S_g + Z)^2] + (E[S_g] + E[Z] ) E[\bar{\bar{S}}], \label{eqn_fourth}
\end{align}
where
\begin{align}
E[\bar{S}] = pE[S_g] + (1-p)E[S_b], \label{eqn_secondlast} \\
E[\bar{\bar{S}}] = qE[S_b] + (1-q)E[S_g]. \label{eqn_last}
\end{align}
In addition, from (\ref{updatecycle_pair}) we have,
\begin{align}
E [Y_b] = E[S_b] + E[Z], \label{eqn_first} \\
E [Y_g] = E[S_g] + E[Z], \label{eqn_second}
\end{align}
since $Y = S+Z$ as defined earlier. 

Substituting (\ref{eqn_third})-(\ref{eqn_second}) in (\ref{avg_age2}) yields the average age expression under Gilbert-Elliot service times and i.i.d.~interarrival times. We note that the numerator of (\ref{avg_age2}) has $pq$ terms whereas the denominator of (\ref{avg_age2}) is linear in $p$ and $q$. 

We note that if the server operates only in state $g$ without switching to state $b$, the average age given in (\ref{avg_age2}) becomes
\begin{align}
	\Delta_g = \frac{E[Q_g]}{E[Y_g]} = \frac{1}{\lambda} + \frac{2}{\mu_g} - \frac{1}{\lambda+\mu_g},
\end{align}
which is the result derived in \cite{Costa16} for an M/M/1 queue with blocking. A similar average age, $\Delta_b$, is achieved if the server node only operates in state $b$.    

\subsection{Gilbert-Elliot Interarrival Times and I.i.d.~Service Times}\label{interarrival}
Service times at the server node are i.i.d.~exponential random variables with rate $\mu$ whereas the interarrival times follow a two-state Markov chain characterized by the state transition matrix in (\ref{state_trans}). State changes occur upon successful entry to the server node. Let $Z_g$ denote the waiting time until the next arrival when the interarrival state is $g$, and let $Z_b$ denote the waiting time when the interarrival state is $b$. Thus,
\begin{align}
(Y_j, Q_j) =  \begin{cases} 
(Y_b, Q_b),  & \text{if } Z_j = Z_b \\
(Y_g, Q_g),  & \text{if } Z_j = Z_g.
\end{cases} \label{updatecycle_pair2}
\end{align} 
Here, similar stationarity and ergodicity arguments apply and the average age is again given by
\begin{align}
\Delta = \frac{E[Q]}{E[Y]} = \frac{qE[Q_b] + pE[Q_g]}{qE[Y_b] + pE[Y_g]}. \label{avg_age3}
\end{align}
By inspecting Fig.~\ref{fig:ageEvol}, we find
\begin{align}
E [Q_b] =& \frac{1}{2}E[(S + Z_b)^2] + E[S]^2 + E[S]E[Z_b], \label{eqn_seventh} \\
E [Q_g] =& \frac{1}{2}E[(S + Z_g)^2] + E[S]^2 + E[S]E[Z_g]. \label{eqn_eighth}
\end{align}
In addition, we have,
\begin{align}
E [Y_b] = E[S] + E[Z_b], \label{eqn_fifth} \\
E [Y_g] = E[S] + E[Z_g]. \label{eqn_sixth}
\end{align}
by using (\ref{updatecycle_pair2}) and the fact that $Y = S+Z$.

Substituting (\ref{eqn_seventh})-(\ref{eqn_sixth}) in (\ref{avg_age3}) yields the average age expression under Gilbert-Elliot interarrival times and i.i.d.~service times. We note that both the numerator and denominator of (\ref{avg_age3}) are linear in $p$ and $q$.

So far, we derived average age expressions for Gilbert-Elliot servers and samplers for a given state transition matrix $P$. We optimize this matrix $P$ in the next section to achieve minimum average age at the monitor node in both scenarios. 

\section{Age-Optimal Transition Matrix $P$}\label{age_opt}
In what follows we characterize the age-optimal state transition matrix $P$ for Gilbert-Elliot service times and Gilbert-Elliot interarrival times.
\subsection{Gilbert-Elliot Service Times and I.i.d.~Interarrival Times}\label{age_opt_service}

In Section~\ref{service} the average age expression (\ref{avg_age2}) for given state transition matrix $P$ under Gilbert-Elliot service times is derived. Next two lemmas characterize the behavior of (\ref{avg_age2}) with respect to the state transition probabilities $p$ and $q$.

\begin{lemma}\label{lemma1}
Under Gilbert-Elliot service times, the average age in (\ref{avg_age2}) monotonically decreases in $p$.
\end{lemma}

\begin{Proof}
	To prove the lemma, we take the derivative of (\ref{avg_age2}) with respect to $p$ and show that it is negative. The numerator of the derivative of (\ref{avg_age2}) is
	\begin{align}
		\frac{1}{2}&E[Y_b]E[Y^2_g] \!-\! \frac{1}{2}E[Y_g]E[Y^2_b] \nonumber\\ &+ \!E[Y_b](E[S_g] \!-\! E[S_b])[(1\!-\!q)E[Y_g] \!+\! qE[Y_b]]. \label{cond_p}
	\end{align}
	In (\ref{cond_p}), the last term is already negative as $E[S_g] < E[S_b]$ as stated in Section~\ref{model}. Thus, we need to show that $E[Y_b]E[Y^2_g] - E[Y_g]E[Y^2_b] < 0$. This is indeed true for exponential interarrival times with rate $\lambda$ and exponential service times $S_g$ with rate $\mu_g$ in state $g$ and $S_b$ with rate $\mu_b$ in state $b$ where $\mu_g > \mu_b$. With this, the result follows.  
\end{Proof}

Thus, as $p$ increases, a better age performance is achieved at the monitor node. This is an intuitive result as larger $p$ indicates that the service state spends more time in the good state $g$ as implied by (\ref{stat_dist}). We note that this result does not depend on $q$ and is valid for any $q \in (0,1)$. 

\begin{lemma}\label{lemma2}
	Under Gilbert-Elliot service times, the average age in (\ref{avg_age2}) monotonically increases in $q$.
\end{lemma}

The proof of Lemma~\ref{lemma2} follows similarly to that of Lemma~\ref{lemma1}. Thus, as $q$ decreases, a better age performance is achieved at the monitor node. Similar to Lemma~\ref{lemma1}, Lemma~\ref{lemma2} holds true for any $p$ value in $(0,1)$. 

From Lemmas~\ref{lemma1} and \ref{lemma2}, we observe that, to achieve the minimum average age under Gilbert-Elliot service times, we need to maximize the time spent in state $g$. 
 
\subsection{Gilbert-Elliot Interarrival Times and I.i.d.~Service Times}\label{age_opt_arrival}

In Section~\ref{interarrival} the average age expression (\ref{avg_age3}) for given state transition matrix $P$ under Gilbert-Elliot interarrival times is derived. Next two lemmas, which follow similar to Lemmas~\ref{lemma1} and~\ref{lemma2}, characterize the behavior of (\ref{avg_age3}) with respect to the state transition probabilities $p$ and $q$.

\begin{lemma}\label{lemma3}
	Under Gilbert-Elliot interarrival times, the average age in (\ref{avg_age3}) monotonically decreases in $p$.
\end{lemma}

Thus, as $p$ increases, a better age performance is achieved at the monitor node as in Gilbert-Elliot service times scenario.

\begin{lemma}\label{lemma4}
	Under Gilbert-Elliot interarrival times, the average age in (\ref{avg_age3}) monotonically increases in $q$.
\end{lemma}

Thus, as $q$ decreases, a better age performance is achieved at the monitor node as in Gilbert-Elliot service times scenario. 

From Lemmas~\ref{lemma3} and \ref{lemma4}, we observe that, to achieve the minimum average age under Gilbert-Elliot interarrival times, we need to maximize the time spent in state $g$. Although more frequent sampling may overwhelm the network and incur higher age in FCFS queueing systems as shown in \cite{Kaul12a}, in our model, since a dropping policy is implemented, more frequent sampling is desirable to obtain lower average age. 

In the next section, we find the age-optimal state transition matrix $P$ when there is an average cost constraint.

\section{Age-Optimal Transition Matrix $P$ under Average Cost Constraint}
\begin{figure}[t]
	\begin{center}
		\subfigure[\label{c_2}]{%
			\includegraphics[width=0.45\linewidth]{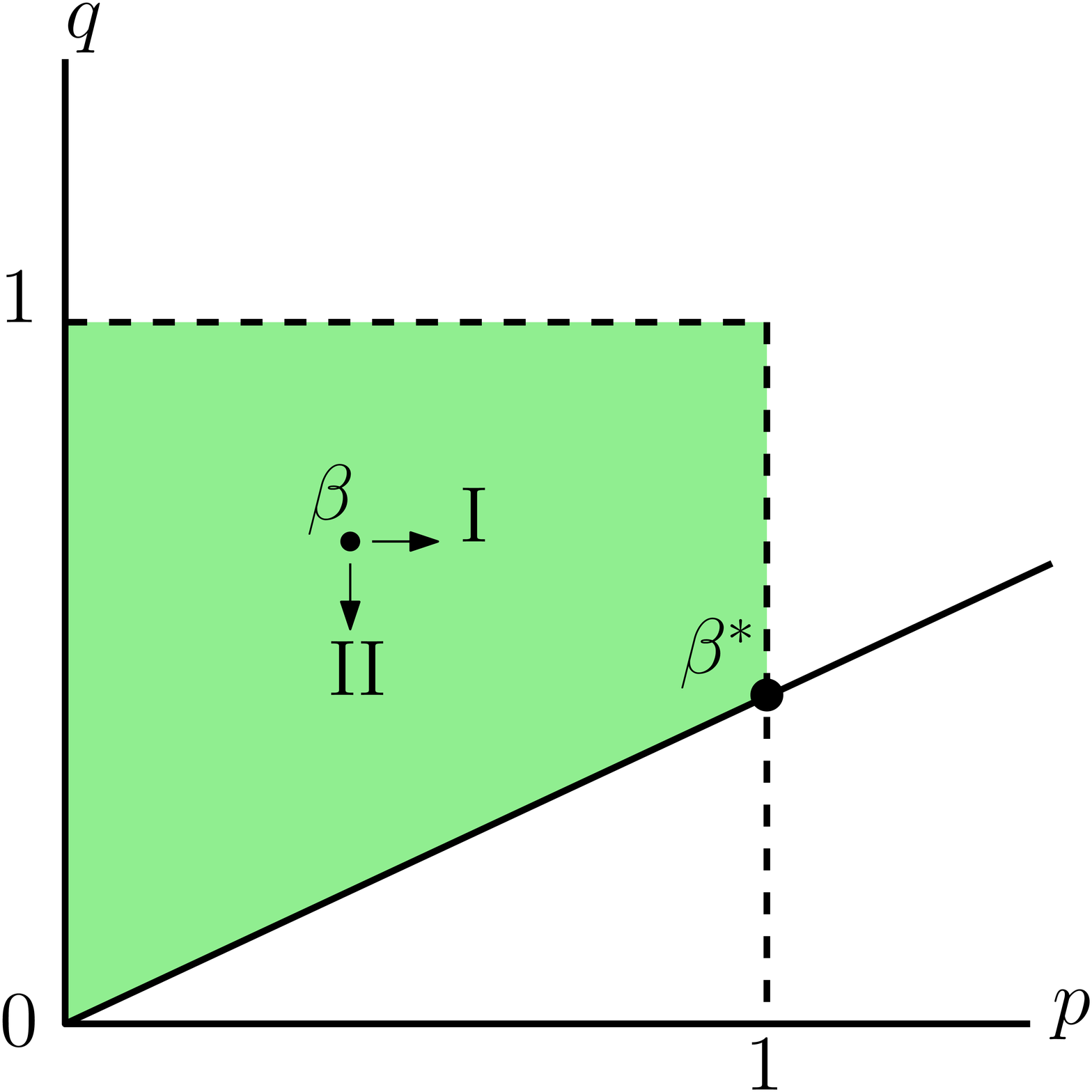}}\vspace{-3mm}
		\subfigure[\label{c3}]{%
			\includegraphics[width=0.45\linewidth]{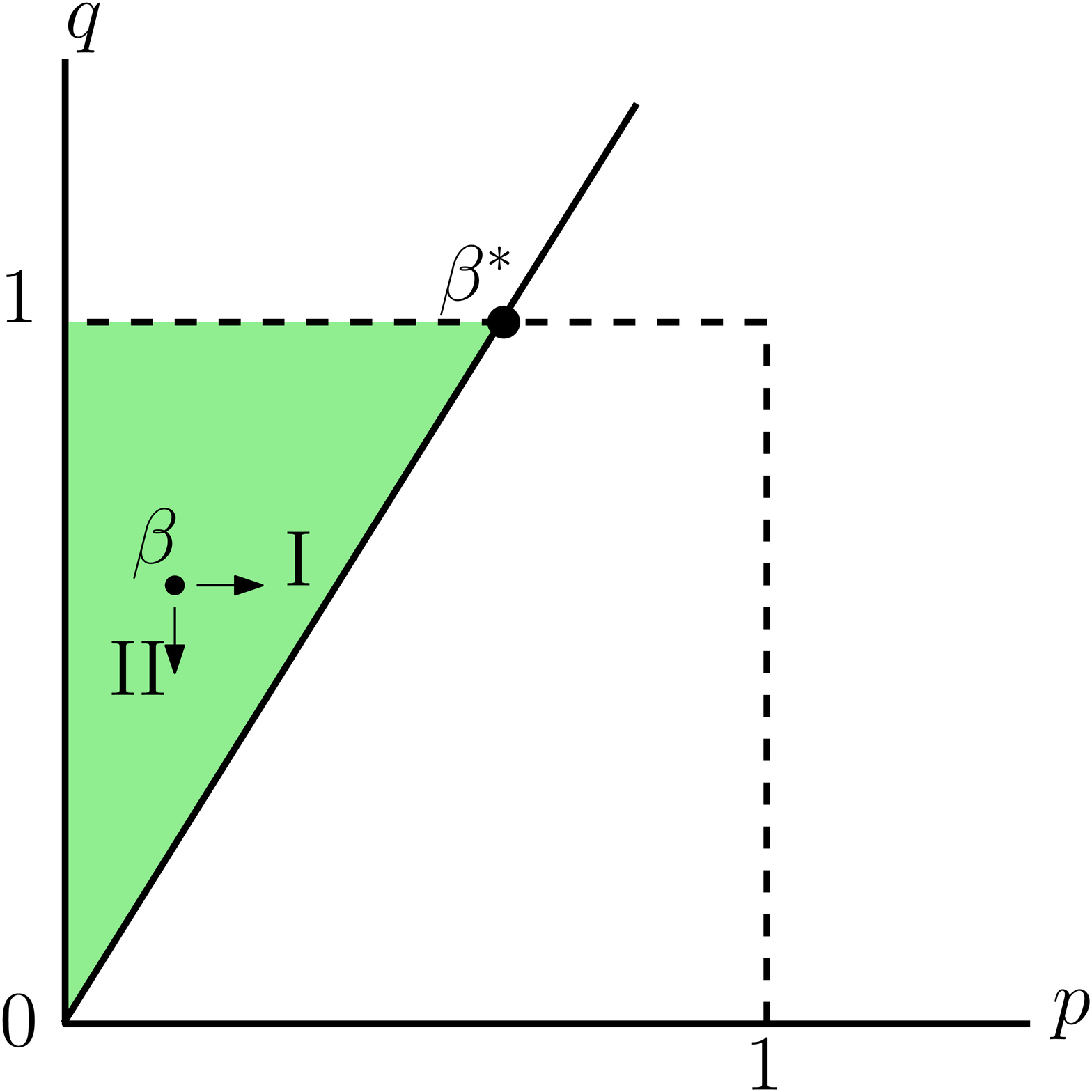}}
		\caption{Feasible $(p,q)$ pairs for the problem in (\ref{problem1}) when (a) $\frac{c-c_b}{c_g-c} > 1$ and (b) $\frac{c-c_b}{c_g-c} < 1$ where the line is $q(c_b-c) + p(c_g-c) = 0$ in both cases.}
		\label{fig:feasible}
	\end{center}
	\vspace{-5mm}
\end{figure}

In Section~\ref{age_opt}, the age optimization is over all possible $p$ and $q$ pairs, i.e., $(p, q) \in (0,1) \times (0,1)$ and we showed that as $p \rightarrow 1$ and $q \rightarrow 0$, the minimum age is achieved in both scenarios. However, when there is a constraint on the operation of the system, all of $(0,1) \times (0,1)$ region may not be feasible. To explore the age-optimal state transition probabilities in such a scenario, here, we consider a constraint on the average cost which may correspond possibly to limited energy budget for the system. Let $c_b$ and $c_g$ denote the cost of operating in state $b$ and state $g$, respectively, where $c_g \geq c_b$ as faster operation requires higher cost (e.g., more energy). When the overall budget is $c$ units, we need to satisfy
\begin{align}
\frac{q}{p+q}c_b + \frac{p}{p+q}c_g \leq c. \label{const1}
\end{align}
Then, the problem to solve becomes,
\begin{align}
\label{problem1}
\min_{\{p, q \}}  \quad &  \frac{qE[Q_b] + pE[Q_g]}{qE[Y_b] + pE[Y_g]} \nonumber \\
\mbox{s.t.} \quad &q(c_b-c) + p(c_g-c) \leq 0, 
\end{align}
where expectations are as in (\ref{eqn_third})-(\ref{eqn_second}) for Gilbert-Elliot service times and as in (\ref{eqn_seventh})-(\ref{eqn_sixth}) for Gilbert-Elliot interarrival times and the constraint follows from (\ref{const1}). The trivial case is when $c \geq c_g \geq c_b$ for which the feasible region is $(0,1) \times (0,1)$ and the results from Section~\ref{age_opt} apply. Thus, in this section, we consider $c_g \geq c \geq c_b$ for which the feasible set is shown in Fig.~\ref{fig:feasible}.
Next, we show that the constraint in (\ref{problem1}) needs to be satisfied with equality.
\begin{lemma}
	Age-optimal $(p,q)$ pair satisfies the constraint in (\ref{problem1}) with equality, i.e., $q(c_b-c) + p(c_g-c) = 0$.
\end{lemma}
\begin{Proof}
	Given a point $\beta$ in the feasible set as shown in Fig.~\ref{fig:feasible}, a lower average age can be obtained as we move along direction I to increase $p$ or along direction II to decrease $q$ as shown in Lemmas~\ref{lemma1} and~\ref{lemma2} under Gilbert-Elliot service times and in Lemmas~\ref{lemma3} and~\ref{lemma4} under Gilbert-Elliot interarrival times. Thus, no point that is not along the $q(c_b-c) + p(c_g-c) = 0$ line can be optimal as we can achieve a lower average age by moving towards this line. 
\end{Proof}

Thus, the age-optimal $(p,q)$ pair is such that $q = \alpha p$ where $\alpha = \frac{c_g-c}{c-c_b}$. With this, the problem in (\ref{problem1}) reduces to a minimization over  probability $p$ only. That is, the objective function in (\ref{problem1}) becomes
\begin{align}
\Delta(p) = \frac{\alpha E[Q_b] + E[Q_g]}{\alpha E[Y_b] + E[Y_g]}, \label{delta_p}
\end{align}
where $\alpha$ is fixed. Next, we solve this problem for Gilbert-Elliot service times and Gilbert-Elliot interarrival times and find the age-optimal $(p,q)$ pair that minimizes the average age.

\subsection{Gilbert-Elliot Service Times and I.i.d.~Interarrival Times}\label{age_opt_service_cost}

We note that, under Gilbert-Elliot service times, the denominator of (\ref{delta_p}) does not depend on $p$ whereas the numerator depends on $p$ linearly. One can show that (\ref{delta_p}) is a decreasing function of $p$ with an argument similar to that of Lemma~\ref{lemma1}. Thus, provided that $\alpha < 1$, the age-optimal transition matrix $P$ under average cost constraint is such that $p\rightarrow 1$ and $q \rightarrow \alpha$ which is the $\beta^*$ point in Fig.~\ref{fig:feasible}(a). This result is intuitive as it maximizes the transition probability from state $b$ to state $g$ and spends fraction of time in state $b$ to satisfy the budget requirement. For example, when $c_g = 2$, $c=1.8$ and $c_b = 1$ we find $\alpha = \frac{1}{4}$ and the optimal selection is $p\rightarrow 1$ and $q \rightarrow \frac{1}{4}$. 

Otherwise, when  $\alpha > 1$, the age-optimal selection is $p \rightarrow \frac{1}{\alpha}$ and $q \rightarrow 1$  which is the $\beta^*$ point in Fig.~\ref{fig:feasible}(b). This result tells us that in the age minimizing operation the transition probability from state $g$ to state $b$ approaches $1$ since the transition probability from state $b$ to state $g$ is already limited by $\frac{1}{\alpha}$. For example, when $c_g = 2$, $c=1.2$ and $c_b = 1$ we find $\alpha = 4$ and the age-optimal selection is $p\rightarrow \frac{1}{4}$ and $q \rightarrow 1$. 

\subsection{Gilbert-Elliot Interarrival Times and I.i.d.~Service Times}\label{age_opt_arrival_cost}

We note that, under Gilbert-Elliot interarrival times, both the numerator and the denominator of (\ref{delta_p}) do not depend on $p$. Thus, any $p$ and corresponding $q = \alpha p$ for the given $\alpha$ yields the same average age. In other words, as long as we operate in $(0, \beta^*)$ in Fig.~\ref{fig:feasible}, i.e., satisfy (\ref{const1}) with equality, we obtain the optimal average age since (\ref{avg_age3}) depends on $p$ and $q$ only through the stationary probabilities of the states given in (\ref{stat_dist}) as expectations in (\ref{eqn_seventh})-(\ref{eqn_sixth}) do not depend on $p$ and $q$.

\begin{figure}[t]
	\centering  \includegraphics[width=0.85\columnwidth]{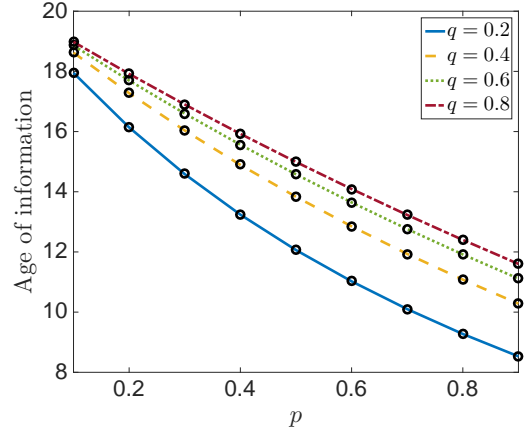}
	\caption{Age of information as a function of probability $p$ for Gilbert-Elliot service times. Symbol $\circ$ marks the simulation results and curves indicate the values obtained from (\ref{avg_age2}). }
	\label{fig:agevsp}
	\vspace{-0.5cm}
\end{figure}
\begin{figure}[t]
	\centering  \includegraphics[width=0.85\columnwidth]{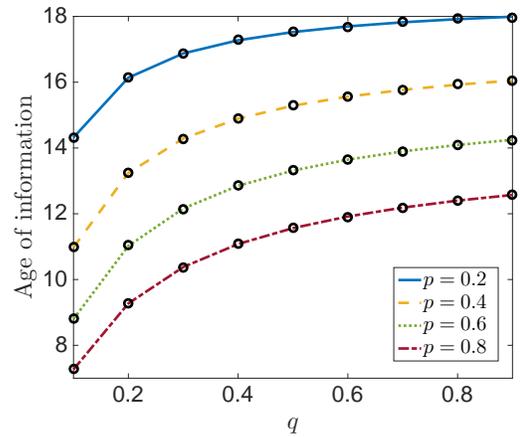}
	\caption{Age of information as a function of probability $q$ for Gilbert-Elliot service times. Symbol $\circ$ marks the simulation results and curves indicate the values obtained from (\ref{avg_age2}). }
	\label{fig:agevsq}
	\vspace{-0.5cm}
\end{figure}

\section{Numerical Results} \label{num_results}

In this section, we provide simple numerical results to validate our theoretical results for a system with arbitrary exponential interarrival and service times. 

We consider Gilbert-Elliot service times and take $\lambda=1$ which is the rate of Poisson arrivals to the server node. We model the service times with an exponential random variable with rate $\mu_b = 0.1$ in state $b$ and with rate $\mu_g = 1$ in state $g$. In Figs.~\ref{fig:agevsp} and~\ref{fig:agevsq}, we plot the average age of information under Gilbert-Elliot service times as a function of the state transition probabilities $p$ and $q$, respectively. In both of the figures, we plot simulation results, marked with $\circ$ symbol, along with results obtained from (\ref{avg_age2}) and observe that the results match. Fig.~\ref{fig:agevsp} shows that the average age decreases monotonically as probability $p$ gets larger as shown in Lemma~\ref{lemma1}. Here, we also note that as $q$ gets lower for a fixed $p$ value, we achieve a lower average age as discussed earlier in Section~\ref{age_opt}. Fig.~\ref{fig:agevsq} shows that the age of information increases monotonically as probability $q$ increases. We also observe that for a fixed $q$ value, the best age is obtained when the $p$ probability is the largest. When the server node only operates in the good state $g$, we find that $\Delta = 2.5$. We observe that this value is lower than the age values in Figs.~\ref{fig:agevsp} and~\ref{fig:agevsq} as the server does not slow down by switching to the bad state $b$. Similarly, if the server node only operates in the bad state $b$, we find that $\Delta = 20.09$ which is strictly larger than the age values shown in  Figs.~\ref{fig:agevsp} and~\ref{fig:agevsq}. Thus, in this case, the server node benefits from switching to the good state $g$.
 
\section{Conclusions}

In this work, we considered an information update system in which status update packets are generated by a sampler and sent to a monitor node through a server node. We considered two scenarios: Gilbert-Elliot service times and i.i.d.~interarrival times; and Gilbert-Elliot interarrival times and i.i.d. service times. In these scenarios, either the server or the sampler follows a two-state Markov chain with the good state $g$ and the bad state $b$ where the operation is faster in state $g$. We determined the average age at the monitor node for both scenarios and characterized the age-optimal state transition matrix for the underlying Markov chain with and without an average cost constraint on the operation of the system.

\bibliographystyle{unsrt}
\bibliography{IEEEabrv,lib_v5}

\end{document}